\documentstyle[12pt,twoside,psfig]{article}
\begin{document}

\begin{center}
{\Large\bf  An interacting scalar field and the recent cosmic acceleration}
\\[15mm]
Sudipta Das \footnote{E-mail: sudipta\_123@yahoo.com}  ~and~  
Narayan Banerjee \footnote{E-mail: narayan@juphys.ernet.in}

{\em Relativity and Cosmology Research Centre,\\Department of Physics, Jadavpur
University,\\Kolkata - 700 032, ~India.}\\
\end{center}

\vspace{0.5cm}
\vspace{0.5cm}

\pagestyle{myheadings}
\newcommand{\be}{\begin{equation}}
\newcommand{\ee}{\end{equation}}
\newcommand{\bea}{\begin{eqnarray}}
\newcommand{\eea}{\end{eqnarray}}
\def\atridot{\stackrel{...}{a}}

\begin{abstract}
In this paper it is shown that the Brans - Dicke scalar field itself can 
serve the purpose of providing an early deceleration and a late time 
acceleration 
of the universe without any need of quintessence field if one considers 
an interaction, i.e, transfer of energy between the dark matter 
and the Brans - Dicke scalar field.
\end{abstract}

\section{Introduction}
Over the last few years, the speculation that our universe is undergoing an 
accelerated expansion has turned into a conviction. The recent observations 
regarding the luminosity - redshift relation of type Ia supernovae \cite{riess} 
and also the observations on Cosmic Microwave Background Radiation (CMBR) 
\cite{lange}  very strongly indicate this acceleration. 
These observations naturally 
lead to the search for some kind of matter field which would generate 
sufficient negative pressure to drive the present acceleration. 
Furthermore, observations reveal that this unknown form of matter, popularly 
referred to as the ``dark energy", accounts for almost 70\% of the present 
energy of the universe. This is confirmed by the very recent Wilkinson 
Microwave Anisotropy Probe (WMAP) data \cite{bridle}. 
A large number of possible 
candidates for this ``dark energy" component has already been proposed and 
their behaviour have been studied extensively. There are excellent reviews 
on this topic \cite{aas}. 
\par It deserves mention that this alleged acceleration should only be a very 
recent phenomenon and the universe must have undergone a deceleration 
(deceleration parameter $q = -\frac{\ddot{a}/a}{\dot{a}^2/a^2} > 0$) in the 
early phase of matter dominated era. This is crucial for the successful 
nucleosynthesis as well as for the structure formation of the universe. 
There are observational evidences too that beyond a certain value of the 
redshift $z$ ( $z \sim 1.5$ ), the universe surely had a decelerated phase 
of expansion \cite{ag}. So, the dark energy component should have evolved 
in such a way that its effect on the dynamics of the universe is dominant 
only during later stages of the matter dominated epoch. A recent work by 
Padmanabhan and Roy Choudhury \cite{paddy} shows that in view of the error 
bars in the observations, this signature flip in $q$ is essential for the 
conclusion that the present universe is accelerating.

\par So, we are very much in need of some form of a field as the candidate for 
dark energy, which should govern the dynamics of the universe in such a 
way that the deceleration parameter $q$ 
was positive in the early phases of the matter dominated era and becomes 
negative during the later stages of evolution. One of the favoured 
choices for the ``dark energy" component is a scalar field called 
a quintessence field ( Q-field ) which slowly rolls 
down its potential such that 
the potential term dominates over the kinetic term 
and thus generates sufficient 
negative pressure for driving the acceleration. A large number of quintessence 
potentials have appeared in the literature and their behaviour have been 
studied extensively ( for a comprehensive review, see \cite{vs} ). However, 
most of the quintessence potentials do not have a proper physical background 
explaining their genesis. In the absence of a proper theoretical plea for 
introducing a particular Q-field, non-minimally coupled 
scalar field theories become attractive for carrying out 
the possible role of the driver of the late time 
acceleration. The reason is simple; the required scalar field is already there 
in the purview of the theory and does not need to be put in by hand. 
Brans - Dicke theory is arguably the most natural choice as the 
scalar - tensor generalization of general relativity (GR) 
because of its simplicity 
and a possible reduction to GR in some limit. 
Obviously Brans - Dicke (BD) theory or its 
modifications have already found some attention as a 
driver of the present cosmic acceleration \cite{nb} (see also \cite{onemli}). 
It had 
also been shown that BD theory can potentially 
generate sufficient acceleration in the matter dominated era even without any 
help from an exotic Q - field \cite{pavon}. 
But this has problems with the required `transition' from 
a decelerated to an accelerated phase. Amongst other nonminimally 
coupled theories, a dilatonic scalar field had also been considered as the 
driver of the present acceleration \cite{piazza}.
\par In most of the models the dark energy and dark matter components are 
considered to be non-interacting and are allowed to evolve independently. 
However, as the nature of these components are not completely known, the 
interaction between them will indeed provide a more general 
framework to work in. 
Recently, Zimdahl and Pavon \cite{zimdahl} have shown 
that the interaction between 
dark energy and dark matter can be very useful in solving the coincidence 
problem ( see also ref \cite{soma} ). Following this idea, 
we consider an interaction 
or `transfer of energy' between the Brans - Dicke scalar field which is a 
geometrical field and the dark matter. The idea of using a `transfer' of 
energy between matter and the nonminimally coupled field had been used earlier 
by Amendola \cite{amen}. We do it specifically for a modified Brans - Dicke 
theory.
The motivation for introducing this 
modification of Brans - Dicke theory is the following. In the presence 
of  matter and a quintessence field, with or without an interaction between 
them, the evolution of net equation of state parameter $w$ plays a 
crucial role in driving a late surge of accelerated expansion. But WMAP 
survey indicates that the time variation of $w$ may be very severely 
restricted \cite{jbp}. If the late acceleration is driven by an exchange of 
energy between matter and a geometrical field $\phi$, the question of 
the variation of $w$ would not arise.

\par We write down the Brans - Dicke field equations in the so called Einstein 
frame. The field equations in this version look simpler and $G$ becomes a 
constant. But one has to sacrifice the equivalence principle as the rest mass 
of a test particle becomes a function of the scalar field \cite{dicke}. So, the 
geodesic equation is no longer valid and the different physical quantities 
loose their significance. Nevertheless, the equations in this version of 
the theory enables us to identify the energy contributions from different 
components of matter. However, for final conclusions we go back to the 
original atomic units where we can talk about the features with confidence. 
We choose a particular form of the interaction and show that 
a constant BD parameter 
$\omega$ can not give us the required flip from a positive to a negative 
signature of $q$ in the matter dominated era. We attempt 
to sort out this problem using a modified form of BD theory 
where $\omega$ is a function of the scalar field $\phi$ \cite{kn}. It has 
been pointed out by  Bartolo and Pietroni \cite{bartolo} 
that a varying $\omega$ 
theory can indeed explain the late time behaviour of the universe.
 By choosing a particular functional form of $\omega$, we show that 
in the interacting scenario, one can obtain a scale factor `$a$' 
in the original version ( i.e, in atomic units ) of the theory so that the 
deceleration parameter $q$ has the desired property of a signature flip 
without having to invoke any quintessence field in the model. We also 
calculate the statefinder pair  \{r,s\}, recently 
introduced by Sahni et al
\cite{alam}, for this model. The statefinder probes the expansion dynamics of 
the universe in terms of higher derivatives of the scale factor, i.e, 
$\ddot{a}$ and $\atridot$. These statefinder parameters along with the SNAP 
data can provide an excellent diagnostic for describing the properties of 
dark energy component in future.

\section{Field Equations and Solutions}
 The field equations for a spatially flat Robertson - Walker spacetime in 
Brans - Dicke theory are 
\bea
3\frac{\dot{a}^2}{a^2} = \frac{\rho_{m}}{\phi} 
      + \frac{\omega}{2}\frac{\dot{\phi}^2}{\phi^2} - 3\frac{\dot{a}}{a}
                 \frac{\dot{\phi}}{\phi}~,~~~~~\\
2\frac{\ddot{a}}{a} + \frac{\dot{a}^2}{a^2} = -\frac{\omega}{2}
       \frac{\dot{\phi}^2}{\phi^2} - \frac{\ddot{\phi}}{\phi} - 
              2\frac{\dot{a}}{a}\frac{\dot{\phi}}{\phi}~.
\eea

The field equations have been written with the assumption that at the present 
epoch the universe is filled with pressureless dust, i.e, $p_{m} = 0$. 
Here $\rho_{m}$ is the matter density of the universe, $\phi$ is the 
Brans - Dicke scalar field, $a$ is the scale factor of the universe and 
$\omega$ is the BD parameter. An overhead dot represents a 
differentiation with respect to time $t$.
\par The usual matter conservation equation has the form 
\begin{center}
$\dot{\rho_{m}} + 3 H \rho_{m} = 0$ ~.
\end{center}
But here we consider an interaction between dark matter and the geometrical 
scalar field and write down the matter conservation equation in the form 
\be 
\dot{\rho_{m}} + 3 H \rho_{m} = Q~~,
\ee
such that the matter field grows or decays at the expense of the 
BD field. The matter itself 
is not conserved here and the nature of interaction is determined by the 
functional form of $Q$. We do not use the wave equation for the BD field 
here because if we treat equations (1), (2) and (3) as independent equations, 
then the wave equation comes out automatically as a consequence of the Bianchi 
identity. It deserves mention that the wave equation will be modified to 
contain $Q$ which will determine the rate of pumping energy from the BD field 
to matter or vice-versa. This interaction term $Q$ is indeed a modification of 
Brans - Dicke theory. But this interaction does not demand any nonminimal 
coupling between matter and the scalar field $\phi$ and hence does not 
infringe the geodesic equation in anyway. In this interaction, the rest 
mass of a test particle is not modified but rather a ``creation'' of matter 
at the expense of the scalar field $\phi$ (or the reverse) takes place. 
In a sense, it has some similarity with the ``C - field'' of the steady 
state theory \cite{hn}.  
\par In the Brans - Dicke theory, the effective gravitational constant is 
given by $G = \frac{G_{0}}{\phi}$, which is indeed not a constant. Now, we 
effect a conformal transformation 
\begin{center}
$\bar{g}_{\mu\nu} = \phi g_{\mu\nu}$~.
\end{center}
In the transformed version $G$ becomes a constant. However, this 
transformation has some limitations which have been mentioned earlier. But 
the resulting field equations look more tractable.
Equations (1) and (2) in the new frame look like
\bea
3\frac{\dot{\bar{a}}^2}{\bar{a}^2} = \bar{\rho} + \frac{(2\omega + 3)}{4}
                        \dot{\psi}^2~,\\
2\frac{\ddot{\bar{a}}}{\bar{a}} + \frac{\dot{\bar{a}}^2}{\bar{a}^2} = 
                   - \frac{(2\omega + 3)}{4} \dot{\psi}^2~,
\eea
and the matter conservation equation takes the form
\be
\dot{\bar{\rho}}_{m} + 3~\frac{\dot{\bar{a}}}{\bar{a}}~\bar{ \rho}_{m} 
                                = \bar{Q}~~,
\ee
where an overbar represents quantities in new frame and $\psi = ln\phi$.
The scale factor and the matter density in the present version are related 
to those in the original version as 
\be 
\bar{a}^2 = \phi a^2~~~and ~~~\rho_{m} = \phi^2~\bar{\rho}_{m}~~.
\ee
Now, we choose the interaction $\bar{Q}$ of the form 
\be
\bar{Q} = -\alpha~\bar{H} \bar{\rho}_{m}~,
\ee
where $\alpha$ is a positive constant. This negative $\bar{Q}$ indicates a 
transfer of energy from the dark-matter (DM) component to the geometrical 
field $\phi$.\\
Equation (6) can be easily integrated with the help of equation (8) to yield
\be
\bar{\rho}_{m} = \rho_{0}~\bar{a}^{(-\alpha - 3)}~,
\ee
where $\rho_{0}$ is a constant of integration.\\
Then, equations (4) and (5) alongwith equation (9) has a solution  
\be
\bar{a} = A \bar{t}^{2/(3 + \alpha)}
\ee
where $A$ is a constant given by
\begin{center}
$A = [\sqrt{\frac{\rho_{0}}{3 - \alpha}}(\frac{3 + \alpha}{2})]^
              {\frac{2}{3 + \alpha}}$~.
\end{center}
Some arbitrary constants of integration have been put equal to zero 
while arriving at equation (10) for the sake of simplicity.\\
Using equations (5) and (10), one can easily arrive at the relation
\be
(\frac{2\omega + 3}{4})\dot{\psi}^2 = \frac{4\alpha}{(3 + \alpha)^2}
                        \frac{1}{\bar{t}^2}~.
\ee
\par If we consider a non-varying $\omega$, equation (11) will give rise to a 
simple power law evolution of $\phi$. From equations (7) and (10), the scale 
factor $a$ in atomic units will also have a power law evolution - 
an ever accelerating 
or an ever decelerating model contrary to our requirement. This is consistent 
with the exhaustive solutions in Brans - Dicke cosmology obtained by 
Gurevich et al. \cite{gurevich}, where the dust solutions are all power law. 
This indicates that the chioce of constants of integration in 
equation (10) does not generically change the model.
One way out of this 
problem is to consider a generalization of Brans - Dicke theory where the 
parameter $\omega$ is a function 
of the scalar field $\phi$ rather than a constant \cite{kn}. An evolving 
$\omega$ will be a contributory factor in determining the dynamics of the 
universe.
\par We make a choice of $\omega$ as, 
\be
\frac{2\omega + 3}{4} = \frac{\alpha}{(3 + \alpha)^2} 
                     \frac{\phi}{(\sqrt{\phi} - 1)^2}~.
\ee
Then, equation (11) can be integrated to yield 
\be
\phi = (1 - \phi_{0}\bar{t})^2~,
\ee
$\phi_{0}$ being a positive constant.
\par It deserves mention here that since $g_{00}$ and $\bar{g}_{00}$ are both 
 equal to one, the time variable transforms as 
\begin{center}
${d\bar{t}}^2 = \phi dt^2$~.
\end{center}
This along with equation (13) gives 
\be
\bar{t} = \frac{1}{\phi_{0}} \left[1 - \sqrt{2\phi_{0}(t_{0} - t)}~\right]~,
\ee
which is a monotonically increasing function of $t$ until $t = t_{0}$, 
beyond which the model really does not work. So one can use $\bar{t}$ 
itself as the new cosmic time in the original version of the theory 
without any loss of generality. So, for the sake of convenience, from 
now onwards, we write $t$ in place of $\bar{t}$.
\par We transform the scale factor back to the original 
units by equation (7), so that we 
are armed with the equivalence principle and can talk about the dynamics 
quite confidently. \\
We have, 
\be
a = \frac{\bar{a}}{\sqrt{\phi}} = \frac{A~t^{2/(3 + \alpha)}}{(1 - \phi_{0}t)}~.
\ee

Also, the Hubble parameter and the deceleration parameter $q$ in the 
original version comes out as,
\bea
H = \frac{2}{3 + \alpha}\frac{1}{t} + \frac{\phi_{0}}{1 - \phi_{0}t}~,\\  
q = -1 + \frac{\frac{2}{3 + \alpha}(1 - \phi_{0}t)^2 - \phi_{0}^2t^2}
            {[\frac{2}{3 + \alpha}(1 - \phi_{0}t) + \phi_{0}t]^2}~.
\eea
\\
\par From equations (15) and (16) it is evident that at $t \rightarrow 
\frac{1}{\phi_{0}}$, both $a$ and $H$ blow up together giving a Big Rip. 
However, this rip has a different characteristic than that engineered by 
a normal phantom field. In the latter, $\rho_{m}$ goes to zero but 
$\rho_{DE}$ goes to infinity at the rip. In the present case, however, there 
is no dark energy as such, and the scalar field is a part of geometry and 
hence it is difficult to recognize its contribution to the energy density. 
In the revised version, however, $\frac{2\omega + 3}{4}{\dot{\psi}}^2$ is 
the contribution towards the stress tensor. It turns out that at 
$t \rightarrow \frac{1}{\phi_{0}}$, this contribution remains quite finite. 
So the big rip is brought into being by the interaction, and not by a 
singularity in the stress tensor.
\begin{figure}[!h]
\mbox{\psfig{figure=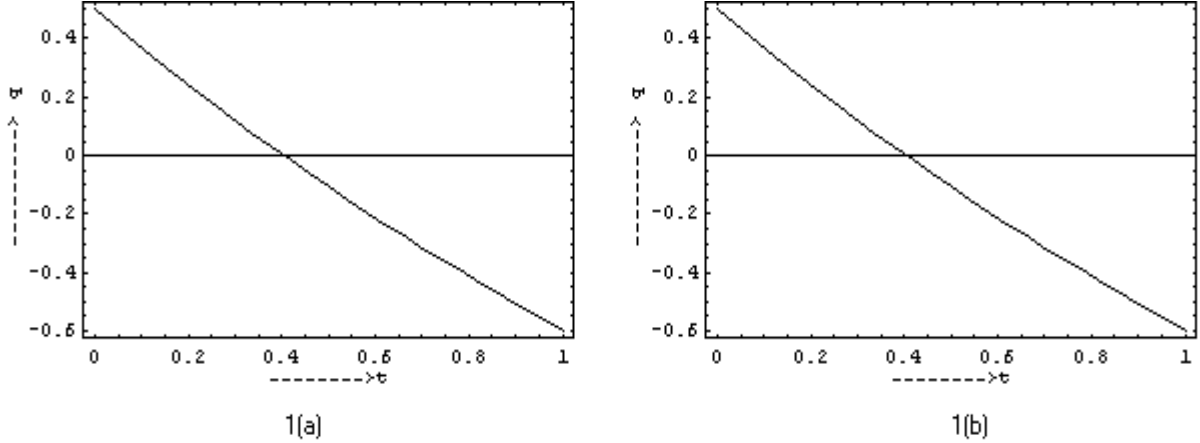,height=2.3in,width=6.333in}}
\caption{Figure 1(a) and 1(b) shows the plot of $q$ vs. $t$ for
different values of $\alpha$. For figure 1(a)
we choose $\alpha = 0.0000003$  whereas for
figure 1(b) we set the value as $\alpha = 0.0000001$.}
\label{first_fig}
\end{figure}
\\
The plot of $q$ against $t$ ( figure 1 ) reveals that the deceleration 
parameter indeed has a sign flip in the desired direction and indicates 
an early deceleration ( $q > 0$ ) followed by a late time acceleration 
( $q < 0$ ) of the universe. Also, the nature of the curve is not crucially 
sensitive to the value of $\alpha$ chosen.
\par Equation (5) clearly indicates that $\ddot{\bar{a}}/\bar{a}$ is 
negative definite. So the signature flip in $q = -\frac{\ddot{a}/a}{\dot{a}^2/a^2}$ in the original Jordan frame must come from the time variation in $\phi$ 
via equation (15). Local astronomical experiments suggest that the 
present variation of $G$ and hence that of $\phi$ has a very stringent 
upper bound. It is 
therefore imperative to check whether the present model is consistent with that 
bound. For figure 1, the value of the constant of integration $\phi_{0}$ is 
fixed at 0.3. With this value, and the age of the universe taken as $\sim$ 
15 Giga years, equation (13) yields
\begin{center}
${\vline~{\frac{\dot{\phi}}{\phi}~\vline}_{~0}}~\sim~10^{-10}$ per year, \\
\end{center}
which is consistent with the requirements of the local experiments \cite{will}. 
The suffix $0$ indicates the present value. Also in this model, from equations 
(12) and (15), we get as $\omega \rightarrow \infty$, $\phi \rightarrow 1$ and 
$a \rightarrow t^{2/(3 + \alpha)}$. Therefore, for very small value of 
$\alpha~ ( \sim 10^{-7})$, $a$ is indistinguishable from that in GR 
$( a \sim t^{2/3})$. This is consistent with the notion that BD theory 
yields GR in the infinite $\omega$ limit.
\section{Statefinder parameters for the model }
\par Recently Sahni et al. \cite{alam} have introduced a pair of new 
cosmological parameters \{r, s\}, termed as ``statefinder 
parameters". These parameters can effectively differentiate between 
different forms of dark energy and provide a simple diagnostic 
regarding whether a particular model fits into the basic observational 
data. These parameters are 
\begin{center}
$r = \frac{\stackrel{...}{a}}{aH^3}$ and 
       $s = \frac{r - 1}{3(q - \frac{1}{2})}$~. 
\end{center}
Accordingly, we find the statefinder parameters for the present model as 
\be
r = 1 + \frac{3\phi_{0}^2 t^2 - 3\beta (1 - \phi_{0}t)^2}
         {[\beta + \phi_{0}t(1 - \beta)]^2} + 
          \frac{2\beta (1 - \phi_{0}t)^3 + 2\phi_{0}^3 t^3}
              {[\beta + \phi_{0}t(1 - \beta)]^3}
\ee
and 
\be
s = \frac{3[\phi_{0}^2 t^2 - \beta(1 - \phi_{0}t)^2]
             [\beta + (1 - \beta)\phi_{0}t] + 2\beta(1 - \phi_{0}t)^3 + 
                       2\phi_{0}^3t^3}
      {3[\beta + (1 - \beta)\phi_{0}t][-\frac{3}{2}{\lbrace \beta + (1 - \beta)
             \phi_{0}t \rbrace} + \beta(1 - \phi_{0}t)^2 - \phi_{0}^2t^2]^2}
\ee
where $\beta = \frac{2}{3 + \alpha}$.
\\
\begin{figure}[!h]
\begin{center}
\mbox{\psfig{figure=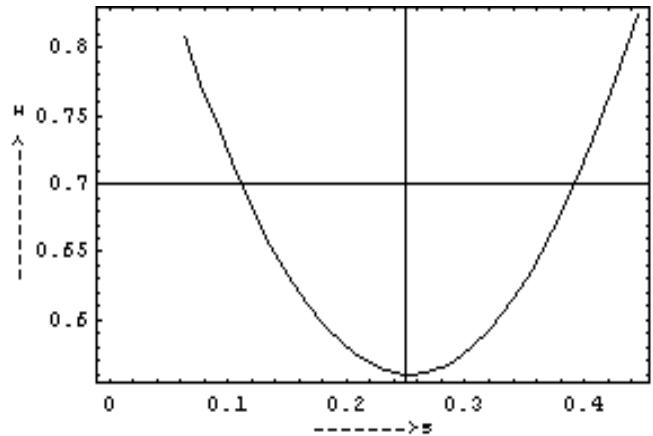,height=2.3in,width=3.5in}}
\caption{Plot of $r$ as a function of $s$ for $\alpha = .0000001$~.}
\label{second_fig}
\end{center}
\end{figure}
\\

\par If we now plot $r(s)$ for some small value of $\alpha$ ( $\alpha << 1$ ), 
we find that the nature of the curve is similar to the one expected for 
scalar field quintessence models with equation of state parameter $w$ in 
the range $-1 < w < 0$ \cite{alam}. 

\section{Discussion}
\par  Thus we see that for a spatially flat universe ($k = 0$), we can 
construct a presently accelerating model in Brans - Dicke theory or more 
precisely in a generalized version of it ( $\omega = \omega(\phi)$ ) 
if one considers an  interaction between 
dark-matter and the geometrical scalar field. It deserves mention that 
for this interaction, the Lagrangian should also be modified by the 
inclusion of an interference term between $\phi$ and $\it{L}_{m}$.
The salient feature of the 
model is that no dark energy component is required. The nature 
of the $q$ vs. $t$ curve 
is not crucially sensitive to small changes in the 
 value of $\alpha$, the parameter which 
determines the strength of the interaction; only the time of `onset' of 
acceleration would change by small amounts with $\alpha$. In revised unit, 
this interaction can be switched off by putting $\alpha = 0$. However, in 
Jordan frame it is not possible to switch off the interaction 
with this particular choice. If we put $\alpha = 0$, the scalar field itself 
becomes trivial. In this frame, the conservation equation ( equation (6) along  
with equation (8) )  transforms to 
\be
\dot{\rho_{m}} + 3H\rho_{m} = - \left[\alpha H + 
               \frac{\alpha - 1}{2}\frac{\dot{\phi}}{\phi} \right]\rho_{m}~.
\ee
As one has both $H\rho_{m}$ and $\frac{\dot{\phi}}{\phi}\rho_{m}$ in the right 
hand side, the transfer of energy between matter and scalar field takes place 
both due to the expansion of the universe and the evolution 
of the scalar field. This equation shows that if 
\begin{center}
$\phi = constant ~ a^{-2\alpha/(\alpha - 1)}$
\end{center}
the interaction vanishes. In this case, the transfer of energy due to 
$H\rho_{m}$ and 
$\frac{\dot{\phi}}{\phi}\rho_{m}$ cancel each other.
\par It is evident from equation (11) that if we consider the interaction 
between dark matter and the geometrical field of the form considered in 
equation (8), a constant $\omega$ will give rise to an ever accelerating 
or ever decelerating model and definitely we are not interested in that. 
So, the idea of varying $\omega$ is crucial here 
as it can very well serve the purpose of providing a signature flip in $q$ 
in this interacting scenario. It deserves mention that the specific 
choice for the interaction in equation (8) and the choice of $\omega = 
\omega(\phi)$ in equation (12) are taken so as to yield the desired result. 
This is indeed a toy model which simply shows that investigations regarding 
an interaction amongst matter and the nonminimally coupled scalar field 
is worthwhile.
\par From equation (11) it is also evident that $(2\omega + 3)$ has to be 
positive definite, i.e, $\omega$ has to pick up some positive value or 
at least $\omega$ should be greater than $-\frac{3}{2}$ in order to sustain a 
consistent model. Also, the parameter $\omega$ does not have any stringent 
limit and thus it may be possible to adjust the value of $\omega$ to some 
higher value. Equation (12) indicates that if $\phi$ is very close to unity, 
which is consistent with the present value of $G$, $\omega$ can attain a high 
value at the present epoch. This is 
encouraging as it might be possible to obtain a model which exhibits early 
deceleration and late time acceleration even with a high value of $\omega$, 
compatible with the limit imposed on it by the 
solar system experiments \cite{will}. 
\par Also, from equation (13) it is evident that $\frac{\dot{\phi}}{\phi} < 0$ 
and $\frac{\ddot{\phi}}{\phi} > 0$. So, from equation (2), which is of 
particular interest in studying the dynamics of the universe, we see that 
the last term $2\frac{\dot{a}}{a}\frac{\dot{\phi}}{\phi}$ is negative and 
is the key factor in driving the present acceleration of the universe. 
This term basically provides the effective negative pressure and becomes 
dominant during the later stages of evolution and drives an accelerated 
expansion. We have also calculated the statefinder parameters for the model 
and show that the \{r, s\} pair mimics that of a quintessence model. 
For this however, the constant $\alpha$ should be given a very small value 
($\sim 10^{-7}$). In this 
model we have considered a particular form of interaction which indeed is not 
unique, and some complicated kind of interaction may lead to more viable 
solutions of the various cosmological problems, particularly to a model 
which is not restricted in future.
\par Although it is true that General Relativity (GR) is by far the best 
theory of gravity and the natural generalisation of BD theory to GR for 
large $\omega$ limit is shown to be restricted in some sense \cite{ss}, still 
BD theory always seems to be ready to provide some useful 
clues to the solution of various 
cosmological problems. The solution to the ``graceful exit" problem 
of inflation 
in terms of an `extended inflation' scenario \cite{la} 
was first obtained in BD theory which provided hints towards the subsequent 
resolutions of the problem in GR. Once again here, BD theory 
in its own right could provide a model 
exhibiting the present cosmic acceleration without introducing any exotic 
dark energy component and since nothing definite is known about the source 
of this acceleration, this type of investigations may lead to some  
track along which viable solutions may finally be arrived at. However, this is 
a primitive model. This only shows that such investigations can be useful. It 
remains to be seen if the solution is an attractor, and whether the model is 
consistent with the structure formation.

\end{document}